# Performance Evaluation of Threshold Signing Schemes in Cryptography

Faneela[1], Jawad Ahmad[2], Baraq Ghaleb[1], Imdad Ullah Khan[3], William J. Buchanan[1], Sana Ullah Jan[1], and Muhammad Shahbaz Khan[1]

[1] School of Computing, Engineering and the Built Environment, Edinburgh Napier University, Edinburgh, United Kingdom.
{F.Faneela, B.Ghaleb, B.Buchanan, S.Jan, muhammadshahbaz.khan}@napier.ac.uk

[2] Cybersecurity Center, Prince Mohammad Bin Fahd University, Alkhobar, Saudi Arabia.
jahmad@pmu.edu.sa

[3] Department of Computer Science, Lahore University of Management Sciences, Lahore, Punjab, Pakistan.
imdad.khan@lums.edu.pk

**Abstract.** Threshold Signature Scheme (TSS) protocols have gained significant attention over the past ten years due to their widespread adoption in cryptocurrencies. The adoption is mainly boosted by Gennaro and Goldfedder's TSS protocol. Since then, various TSS protocols have been introduced with different features, such as security and performance, etc. Large organizations are using TSS protocols to protect many digital assets, such as cryptocurrency. However, the adoption of these TSS protocols requires an understanding of state-of-the-art research in threshold signing. This study describes the holistic view of TSS protocols, evaluates cutting-edge TSS protocols, highlights their characteristics, and compares them in terms of security and performance. The evaluation of these TSS protocols will help the researchers address real-world problems by considering the relevant merits of different TSS protocols.

## 1 Introduction

Digital signatures are the electronic authentication of digital messages and ensure data security. The Digital Signature Scheme (DSS) provides a baseline for developing cryptographic protocols. Different DSS are designed to improve the performance and security of cryptographic protocols. One such development is the threshold signature scheme (TSS), which has gained widespread attention because of its ability to ensure robust data security [8]. TSS is a DSS that allows multiple parties $n$ called signers, to generate a single signature. TSS splits the digital signature into $n$ signature shares or key shares for a message $m$

TSS requires a minimum number, called threshold $t$, of signers $n$ to produce a valid signature to sign a message $m$ using their key shares; this implies that signatures can still be produced even if some of the signers are not active (online), e.g., in the case of $t$ online signers are available. The main feature of TSS is its fault tolerance behavior because an adversary (a malicious user) would need to learn at least $t$ signatures to compromise the message; otherwise, the adversary cannot generate a valid signature. This additional feature of TSS makes it more applicable to real-world scalable systems, such as cryptocurrencies.

TSS has gained significant attention over the last few decades. Desmedt proposed the TSS three decades ago [8], and the proposed scheme peaked after 20-25 years with more advancements in threshold signing such as [3,6]. Blockchain researchers have gathered interest in TSS and incorporated TSS protocols in blockchain using approaches such as [4, 7, 12–14]. To implement TSS in scalable systems, researchers and professionals would need a simplified and updated knowledge of different TSS protocols to improve data security and efficiency. To address this issue, we have evaluated state-of-the-art TSS protocols. TSS protocols are mainly classified into three classes: ECDSA, BLS, and Schnorr-based signatures. However, this study focuses on the TSS protocols using ECDSA and BLS-based signatures, explained in section 1, because these are widely used in industry. We selected the TSS protocols which are frequently evaluated in the literature. In particular, we evaluated the following protocols such as, GG18 [13], GG20 [14], CMP [4], GLOW20 [12] and BLS [18] etc. All of these advanced protocols can be further categorized into their respective classes of TSS signatures. GG18 [13], GG20 [14], CMP [4], GLOW20 [12] are examples of ECDSA-based TSS protocols and BLS [18] is based on BLS-based TSS protocol. The evaluation of TSS protocols is based on security and runtime performance. Security is measured by computational hardness, adversarial structure, and protocol majority, whereas communication rounds are used to evaluate the runtime performance of TSS protocols.

This paper evaluates the state-of-the-art protocols based on TSS, highlights their characteristics, and evaluates them based on security and performance. We evaluated different types of TSS protocols and different open-source TSS libraries based on the most widely used TSS protocols. This would also help professionals make informed decisions about TSS protocols and libraries according to their requirements and deploy them on scalable applications to ensure data security in modern cryptography. For example, the fault tolerance and static adversarial structure of GG18 make them suitable for cryptocurrency applications, where the number of online parties is restricted. Furthermore, the efficiency of GLOW20 due to its one-round structure makes it perfect for IoT devices in smart applications that demand limited resources. The remainder of this paper is organized as follows: Section 1 introduces the basic knowledge required for understanding this study. Section 3 discusses related work. Section 4 presents the evaluation criteria against which different TSS protocols are evaluated. Section 5 theoretically evaluates different TSS protocols. Section 6 shows the empirical evaluation of TSS protocols based on open-source TSS libraries, and finally, Section 7 concludes the study and results.

## 2 Preliminaries

In this section, we present the definition of preliminaries in threshold signing.

### 2.1 Cryptographic Concepts

Cryptography is fundamental to secure communication and data protection. Below are key cryptographic elements and algorithms essential for understanding threshold signing.

- Cryptographic key. A cryptographic key is a set of bit strings used to convert sensitive data into encrypted data using public and private keys. The public key ($pk$) is used to encrypt plain text, while the private key ($sk$) is used to decrypt the encrypted data into plain text [15].
- Pseudorandomness. Pseudorandomness provides a base for generating random numbers for different security algorithms, such as the RSA algorithm.
- Rivest-Shamir-Adleman (RSA) is the oldest cryptographic algorithm used to encrypt and decrypt data. It works similarly to the Digital Signature Algorithm (DSA), a standard for digital signatures, but RSA uses larger cryptographic keys. This key size makes RSA slower for decrypting and signing operations than DSA.
- Decisional Diffie-Hellman (DDH) algorithm. DDH ensures the security of a cryptographic system by making it very hard to decide if two objects are related or different.
- Bilinear pairing method. The bilinear pairing method is used to pair two cryptographic groups to analyze the cryptographic system.
- Elliptical Curve Digital Signature Algorithm (ECDSA). ECDSA uses the Elliptical Curve Cryptography method based on several cryptographic concepts such as Bilinear pairing and DDH etc. ECDSA is used to generate small cryptographic keys, which makes it faster than RSA.

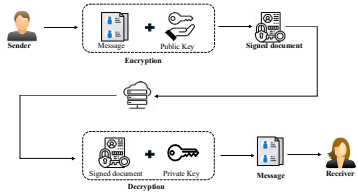

Fig. 1: Basic architecture of digital signature scheme.

### 2.2 Digital Signature Schemes

A digital signature scheme includes three functions: key generation, signing, and signature verification [13], [9]. The key generation function ($Key - gen$) inputs a security parameter $\lambda$ and generates the public key ($sk$) and private key ($pk$) to sign a message $m$. The signing function ($Sig$) takes $sk$ and $m$ and returns a signed message $m_0$, as in equation. After signing a message $m$, the verification function ($Ver$) confirms the signature's validity. It takes $pk$, $m$, and $m_0$ and returns a bit $b$ as indicator. If the signature is valid, the value of $b$ is 1; otherwise, it is 0. Figure 1 shows the basic architecture of a digital signature scheme.

### 2.3   Threshold Signatures

A threshold signature scheme (TSS) is a type of DSS that allows multiple parties to jointly sign a message such that at least $t+1$ out-of-$n$ parties are available for the successful signing of the message, where $t$ is the threshold and $n$ is the total number of parties. Figure 2 provides a simplified view of the TSS. TSS is divided into three functions: threshold key generation, threshold signature, and threshold verification. Threshold key generation function accepts security parameter $\lambda$ as input. Each party in a network receives a share of public and private keys corresponding to this security parameter. In threshold signing $Sig_{(tss)}$, the private key $sk_i$ of $t+1$ of $n$ players are used to sign a message m. The signed message m0 is returned as an output such as:

$$Sig_{(tss)}(sk_i, m) \rightarrow m_0 \qquad (1)$$

Finally, all key shares are combined into one complete secret key. The verification function $Ver_{(tss)}$ uses this key to verify the message. Figure 4 illustrates the detailed process of the threshold signing process.

$$Ver_{(tss)}(pk, sk, m_0) \rightarrow b \qquad (2)$$

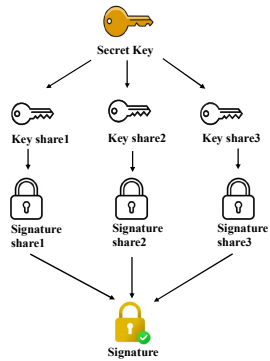

Fig. 2: Simplified view of a threshold signature scheme

## 3   Literature Review

The threshold signature scheme (TSS) enables multiple parties to sign a message using a single signature, reducing network compromise risks. In a recent study, Liu et al. [17] surveyed the threshold signatures in post-quantum cryptography to analyze their strengths in the deployment of TSS. Furthermore, Chen et al. [5] reviewed the lattice-based TSS and classified them according to their functionalities. Based on classical cryptography, TSS is mostly being adopted with ECDSA and BLS signatures.

ECDSA-based TSS is commonly used in elliptical curve cryptography such as the authors in [1] reviewed different TSS protocols based on elliptic curves to highlight their properties for real-world implementations. In [14], the authors introduced an efficient ECDSA protocol for the asynchronous participation of parties. The protocol is intended to identify adversaries who maliciously abort the signing process. Sinan et al. [11] reviewed the issues in the key generation process and recommended a TSS protocol for digital signatures. In [20], the authors proposed a TSS based on time lock puzzles. These puzzles are packaged homomorphically to enhance the security of existing TSS protocols. Furthermore, the authors of [7] introduced a promise $\sum$-protocol for secure and reliable homomorphic operations. The proposed protocol ensured the efficiency of homomorphic operations without using complex cryptographic concepts such as low-order assumptions and zero-knowledge proofs. The authors of [16] highlighted the disadvantages of traditional digital signatures in modern cryptography by emphasizing their time-consuming nature in the verification process.

Moreover, Boneh et al. [3] presented a short signature scheme in TSS called BLS signatures, based on the group theory, bilinear pairing and ECDSA curve. This technique is intended for low-bandwidth, man-typed signatures. In [2], the author proposed a security scheme for a key generation process. The proposed scheme examined the security issues of BLS signatures and demonstrated their vulnerability to adversaries. In a study, [18], the authors presented a Weil pairing scheme and improved the execution time of BLS signatures. Furthermore, Tseng et al. [21] proposed a protocol to identify cheaters during the key signature process. The proposed protocol does not rely on any third party for cheater identification. The above research studies provide effective solutions and references for the TSS. However, there is limited focus on adversarial structures in dynamic and semi-honest settings. Moreover, further investigation is needed to carry out the TSS protocols for real-world applications with increased latency and limited computational resources. This study summarizes the state-of-the-art TSS protocols to help readers understand their strengths and weaknesses for deployment in real-world applications.

## 4   Evaluation Criteria

In this paper, we have evaluated different TSS protocols and libraries theoretically and empirically. In theoretical analysis, different TSS protocols are evalu-ated based on security and runtime performance. However, the empirical evaluation is focused on runtime performance. In the empirical evaluation, we chose two widely used TSS libraries implementing the most commonly used TSS protocols, explained in Section 5.

### 4.1  Security

In this section, we classify the security parameters of the most commonly used TSS protocols into computational hardness, adversarial structures, and protocol majority. These parameters are briefly outlined in the following sections.

Computational Hardness. In cryptography, it is significant to consider the computational hardness of TSS protocols for practical deployment. The computation of TSS protocols is based on some complex mathematical assumptions called computational hardness, also known as the hardness assumption [19]. These mathematical assumptions are complex mathematical problems such as the Elliptical Curve Digital Signature Algorithm (ECDSA), Rivest-Shamir-Adleman (RSA), Diffie-Hellman (DDH) [13], bilinear pairing [18], and pseudorandomness [7], explained in Section 1. In TSS, most of these mathematical problems are extremely difficult and cannot be solved efficiently, which makes it secure against an adversarial attack.

Adversarial Structure. An adversarial structure is an attacking pattern that can successfully intrude on a secure system. In TSS, an adversary may be an algorithm or a sequence of statements that can invade, change, or corrupt the system [10]. There are two primary ways in which an adversary can compromise the security of a TSS protocol: static and dynamic. In a static adversary structure, the adversary decides the targeted parties within a system before the attack occurs. In contrast, an adaptive adversary pattern involves adversaries who do not have a specific target before the execution of the algorithm, and they choose their targeted parties dynamically over time.

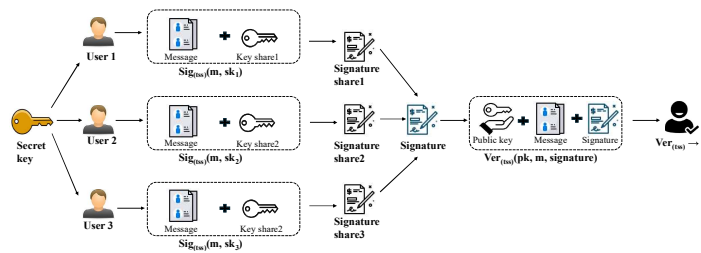

Fig. 3: Detailed process of threshold signing scheme

Protocol Majority. In threshold signing, TSS protocols are designed by assuming the pattern of the majority of involved parties, which is called protocol majority. A party can be either honest or dishonest. A party is assumed to be honest if it is following the rules of protocol. If a party deviates from the rules to cheat, the protocol is said to be dishonest. The protocol majority can be characterized into the honest, dishonest and semi-honest structures. The honest majority is based on the assumption that more than half of the parties are behaving honestly in the TSS protocol, such as when $n/2$ of the $n$ parties are honestly following the protocols. The dishonest majority is based on the assumption that more than half of the parties are behaving dishonestly, such as $n/2$ of the $n$ parties are dishonest in the TSS protocol. The semi-honest majority means that the majority of the parties are honestly following the TSS protocol but are still curious to learn irrelevant information about the other parties.

### 4.2 Runtime Performance: Communication rounds

Communication rounds play an important role in cryptography. The round is iterative step to complete the communication phase with server. Various parameters such as the architecture of the protocol and its hardness assumption affect the number of rounds. Increasing the number of rounds can reduce computational complexity while increasing network latency and processing time. However, reducing the number of rounds can make the network's performance more computationally efficient.

Table 1: Performance comparison of threshold signature protocols

| Protocol | Security | | | Performance | |
|---|---|---|---|---|---|
| | Protocol Majority | Adversarial Structure | Computational Hardness | Key Generation | Signing |
| GG18 [13] | Dishonest | Static | RSA, DDH, ECDSA | 4 | 7 |
| GG20 [14] | Dishonest | Static | RSA, DDH, ECDSA | 4 | 7 |
| Glow20 [12] | Honest | Static | Pseudo-randomness, DDH, ECDSA | 1 | 1 |
| CMP [4] | Dishonest | Dynamic | RSA, DDH, ECDSA | 4 | 5 |
| BLS [18] | Dishonest | Dynamic | Multiplication, Group theory, Bilinear pairing | 3 | 3 |

## 5  Results of Theoretical Evaluation

This section analyzes various state-of-the-art TSS protocols based on the criteria outlined in section 4.

### 5.1  GG18

In 2018, Rosario Gennaro and Steven Goldfeder presented a threshold protocol [13], called GG18. Its security is demonstrated by strong RSA, DDH, and ECDSA with a static dishonest majority. Performance is evaluated based on rounds, both in the key setup and signature phases. The setup phase involves key generation, while the signing phase is the process of key signing and verification. The first stage consists of four rounds, and the second stage consists of seven rounds, including one online round. Moreover, GG18 was the first major development that supported the ECDSA TSS. It has been proven to be more secure because of its foundation on "game-based theory".

### 5.2  GG20

In 2020, Gennaro et. al. presented another protocol [14], GG20, intended for a dishonest majority. The hardness assumption is based on strong RSA, ECDSA, and DDH. It is proposed to deal with static adversaries. In this protocol, performance is evaluated in terms of cryptographic rounds, both in the key setup and signature phases. The setup phase includes the initial setup for key generation. The signing phase encompasses the process of signing a key and verifying that key. The first phase consists of four rounds, and the other stage consists of seven rounds, including one online round. GG20 is an upgraded variant of GG18, with the key development of an identifiable abort. The identifiable abort is a feature to identify the malicious parties terminating the protocol. It is regarded as more efficient because it reduces the number of rounds.

### 5.3  Glow20

David Galindo proposed another protocol in [12], Glow20. It is designed to focus on the honest majority. The hardness of Glow20 is based on DDH and strong pseudo randomness. It is proposed to target the static adversarial structure. Glow20 is a one-round protocol to measure the performance of the key setup and signing phases.

### 5.4  CMP

Ran et al. [4] suggested a protocol, CMP, intended for a dishonest majority. The hardness assumption is based on strong RSA, ECDSA, and DDH. It is designed to deal with the adaptive structure of the adversary. CMP's performance is evaluated in terms of cryptographic rounds, both in the key setup and signature phases. The setup phase includes the initial setup for key generation and key refreshing. The signing phase encompasses the process of signing a key and verifying that key. The first phase consists of five rounds, and the other stage consists of four rounds. Compared to other protocols, the adaptive adversarial behavior of CMP recovers parties corrupted by attackers. Moreover, it corrupts attackers back into a single epoch, where an epoch is the time required for key sharing.

### 5.5 BLS

BLS [18] is a short signature protocol offered by Boneh, Lynn, and Shacham. This short signature scheme is based on the Weil pairing method. Its hardness is based on simple multiplication for key generation and bilinear pairing for the signing method. Moreover, it also uses simple Diffie-Hellman for grouping, such as Gap Diffie-Hellman groups. This is designed for an adaptive dishonest majority. BLS is relatively efficient due to its short signature. Its deployment is generally preferred for applications with low memory requirements.

Table 1 summarizes the performance of different TSS protocols, providing a comparative analysis to guide researchers in selecting the most suitable protocol for their specific real-world applications based on security and runtime performance criteria.

## 6  Empirical Evaluation

In this section, we empirically evaluated the runtime performance of the TSS protocols based on processing time and latency time. Moreover, we also considered another factor, key-threshold $t$, to analyze its impact on runtime performance. All experiments in this study were conducted on a computer equipped with an Intel i7-8550U processor, simulated by manual network growth to monitor the CPU load using processing time and latency time. To evaluate the performance of TSS protocols, we proposed two sets of experiments: processing time and latency time. The experiments are further divided into key generation and key signing phases. To evaluate the performance of TSS protocols, we implemented our experiments on two widely used TSS libraries, MP-ECDSA and MP-BLS. The main reason for choosing these libraries is that they are open-source libraries and easy to configure for researchers. MP-ECDSA is a threshold signature library based on ECDSA signatures [14], available at https://github.com/ZenGo-X/multi-party-ecdsa. MP-BLS is a threshold signature library based on BLS signatures [3] and can be accessed at https://github.com/ZenGo-X/multi-party-bls. For each of these settings, we have run the most widely used TSS protocol, GG20 [14] with different values of $n$ against different sets of threshold $t$. In all these tests, we used key generation and key signing as different phases for the aforementioned TSS libraries. In each of our experiments, we report the processing time and latency time for different sets of $t$ and $n$. However, the values of t were chosen to reflect common settings in research experiments, offering robustness and fault tolerance. Table 2 and Table 3 show the processing time and latency time of both libraries in the key generation and signing process, respectively.

In this setup, we observed that the key-generation process is generally more computationally expensive than the key-signing processes. This is mainly due to the reason that the maximum number of keys need to be stored in the system for the Key-generation process. In contrast, the Key-signing process demands only the keys for minimum active signers, such as $n/2+t$. While this subset can be significantly smaller than the total number of parties $n$, the exact proportion depends on the value of $t$. It is also noticed that the increase in the number of parties $n$ causes a significant rise in processing and latency time, whereas the effect threshold $t$ does not come up with a major change. Importantly, we observed that $MP-ECDSA$ appears to be more efficient than $MP-BLS$ with shorter processing time and latency time. However, the shorter processing time and latency time of $MP-ECDSA$ make it ideal for efficient environments. Similarly, $MP-BLS$ may be employed in high-security scenarios with less efficiency requirements. This is mainly due to the evaluation criteria (explained in Section 4) of TSS protocols used in the libraries, as mentioned earlier. The computational hardness of [14] is based on mathematically complex algorithms that can be reduced by increasing the number of communication rounds. The reduced complexity of computational hardness leads to the static adversarial design, making it suitable for individual signatures with short lengths of signing keys. Similarly, [3] seems to be less efficient than [14]. The reduced number of communication rounds in [3] leads to the high complexity of the bilinear pairing algorithm against dynamic adversaries. This complex structure and strong security feature make it suitable for distributed applications.

## 7   Conclusion

We believe that TSS protocols are core building blocks of modern cryptography. One of the key challenges in their adoption is understanding the deployment of these protocols. In this paper, we first simplified the TSS and evaluated the cutting-edge TSS protocols based on security and performance measures. Our analysis and research classified security into three categories: computational hardness, adversarial structure, and protocol majority. The performance is evaluated based on the number of communication rounds in different processes of TSS protocols. In addition, we evaluated the performance of two different TSS libraries based on the most commonly used TSS protocols. Our evaluation shows that performance and security are interrelated, such that increasing security will cause a decrease in the runtime performance of the TSS protocols. In the future, we would like to extend our evaluation to more advanced adversarial structures in post-quantum cryptography that would help researchers and professionals have a simplified understanding of state-of-the-art TSS protocols to implement in real-world applications.

Table 2: Processing Time of Threshold Signature Libraries

| Threshold ($t$) | Number of Parties ($n$) | Key-Generating Time (s) | | Key-Signing Time (s) | |
|---|---|---|---|---|---|
| | | MP-BLS | MP-ECDSA | MP-BLS | MP-ECDSA |
| 1 | 5  | 5.035    | 0.259  | 1.836  | 1.836  |
|   | 10 | 28.987   | 1.294  | 2.201  | 2.201  |
|   | 15 | 110.187  | 2.895  | 2.881  | 2.881  |
|   | 20 | 251.894  | 5.140  | 4.374  | 4.374  |
|   | 30 | 843.193  | 12.267 | 10.926 | 10.926 |
| 2 | 5  | 5.752    | 0.477  | 1.349  | 1.349  |
|   | 10 | 41.732   | 1.321  | 3.613  | 3.613  |
|   | 15 | 149.897  | 2.880  | 5.306  | 5.306  |
|   | 20 | 364.064  | 5.288  | 6.229  | 6.229  |
|   | 30 | 1208.893 | 12.788 | 11.445 | 11.445 |
| 3 | 5  | 7.352    | 0.320  | 1.151  | 1.151  |
|   | 10 | 61.509   | 1.290  | 2.371  | 2.371  |
|   | 15 | 201.807  | 3.036  | 4.039  | 4.039  |
|   | 20 | 479.516  | 5.654  | 7.581  | 7.581  |
|   | 30 | 1614.058 | 13.091 | 14.513 | 14.513 |

Table 3: Latency Time of Threshold Signature Libraries

| Threshold ($t$) | Number of Parties ($n$) | Key-Generating Time (s) | | Key-Signing Time (s) | |
|---|---|---|---|---|---|
| | | MP-BLS | MP-ECDSA | MP-BLS | MP-ECDSA |
| 1 | 5  | 2.065  | 1.695  | 1.200  | 0.970 |
|   | 10 | 4.243  | 3.688  | 2.882  | 2.096 |
|   | 15 | 6.538  | 5.719  | 4.281  | 2.873 |
|   | 20 | 8.588  | 7.640  | 6.741  | 4.227 |
|   | 30 | 12.711 | 11.644 | 11.500 | 6.809 |
| 2 | 5  | 2.342  | 2.008  | 1.720  | 1.555 |
|   | 10 | 4.548  | 4.445  | 3.583  | 3.041 |
|   | 15 | 6.882  | 6.631  | 4.813  | 3.668 |
|   | 20 | 8.576  | 8.875  | 7.000  | 5.091 |
|   | 30 | 13.167 | 12.612 | 12.778 | 6.959 |
| 3 | 5  | 1.758  | 1.776  | 1.928  | 1.717 |
|   | 10 | 3.789  | 3.693  | 3.816  | 2.765 |
|   | 15 | 6.170  | 5.675  | 5.134  | 3.811 |
|   | 20 | 7.528  | 8.533  | 7.578  | 5.557 |
|   | 30 | 12.961 | 13.932 | 14.729 | 6.122 |